# Hydrogen induced electronic transition within correlated perovskite nickelates with heavy rare-earth composition

*Yi Bian[1†], Haiyan Li[1†], Fengbo Yan[1†], Haifan Li[1], Jiaou Wang[3], Hao Zhang[2], Yong Jiang[1], Nuofu Chen[3] and Jikun Chen[1]\**

[†]Y. Bian, H. Li and F. Yan contribute equally to this work.

[1]School of Materials Science and Engineering, University of Science and Technology Beijing, Beijing 100083, China

[2]School of Renewable Energy, North China Electric Power University, Beijing 102206, China

[3]Beijing Synchrotron Radiation Facility, Institute of High Energy Physics, Chinese Academy of Sciences, Beijing 100049, China

Correspondence: Prof. Jikun Chen (jikunchen@ustb.edu.cn)

.

**Abstract**

Although discovery in hydrogen induced electronic transition within perovskite family of rare-earth nickelate ($Re$NiO$_3$) opens up a new paradigm in exploring both the new materials functionality and device applications, the existing research stays at $Re$NiO$_3$ with light rare-earth compositions. To further extend the cognition towards heavier rare-earth, herein we demonstrate the hydrogen induced electronic transitions for quasi-single crystalline $Re$NiO$_3$/LaAlO$_3$ (001) heterostructures, covering a large variety of the rare-earth composition from Nd to Er. The hydrogen induced elevations in the resistivity of $Re$NiO$_3$ ($R_H/R_0$) show an unexpected non-monotonic tendency with the atomic number of the rare-earth composition, e.g., firstly increase from Nd to Dy and afterwards decreases from Dy to Er. Although $Re$NiO$_3$ with heavy rare-earth composition (e.g. DyNiO$_3$) exhibits large $R_H/R_0$ up to $10^7$, their hydrogen induced electronic transition is not reversible. Further probing the electronic structures via near edge X-ray absorption fine structure analysis clearly demonstrates the respective transition in electronic structures of $Re$NiO$_3$ from Ni$^{3+}$ based electron itinerant orbital configurations towards the Ni$^{2+}$ based electron localized state. Balancing the hydrogen induced transition reversibility with the abruption in the variations of material resistivity, we emphasize that the $Re$NiO$_3$ with middle rare-earth compositions (e.g. Sm) to be most suitable that caters for the potential applications in correlated electronic devices.

The hydrogen induced electronic phase transitions within d-band correlated metal oxides open up a new paradigm to explore new material functionality and device applications via directly manipulating the orbital occupancy and electronic structures[1-3]. As a representative such family of material, the rare-earth nickelates (ReNiO$_3$) exhibits exceptional sensitive electronic structures to hydrogen and experiences multiple electronic transitions via chemical/electrochemical hydrogen doping[1,4-6]. Upon hydrogenations, the typical Mott insulating state of the electron itinerant $t_{2g}^6 e_g^{1\pm\Delta}$ (or $t_{2g}^6 e_g^1$) orbital configuration based on Ni$^{3+}$ transits towards the highly electron localized $Ni^{2+}t_{2g}^6 e_g^2$ state based on Ni$^{2+}$. As a result, the material resistivity of ReNiO$_3$ is abruptly elevated reversibly by several orders of magnitudes, while meanwhile maintains high proton conducting properties. The discovery of such hydrogen triggered electronic transition properties beyond conventional semiconductors recently enables new applications, such as ocean electric field sensing [6], bio-sensing[7], proton-gated electronic devices[1,8,9], neuro-synapse artificial intelligence[4,10], and correlated fuel cells[5].

Nevertheless, the present cognition in the hydrogen induced electronic transitions of ReNiO$_3$ stays at the light rare-earth composition (e.g. Nd, Sm and Eu), from the perspective of both fundamental investigation and device applications[1,4-6,11]. It is worth noticing that the distinguished advantage for ReNiO$_3$ is the high tolerance in designing its electronic structures and properties via the rare-earth composition occupying the A-site of the distorted perovskite structure (ABO$_3$)[12]. For example, reducing the ionic radius of the rare-earth element ($r_{Re}$) or enlarging the atomic number of the rare-earth elements causes a more distorted NiO$_6$ octahedron in crystal structure, and this widens the energy band gap as split from the charge disproportionated Ni$^{3+}$ valance[12-15]. As a result, the relative stability in the insulating or metallic electronic phases can be flexibly regulated via the rare-earth composition, and the respective metal to insulator transition temperature ($T_{MIT}$) is widely adjustable within a broad temperature range of 100-600 K[12,16-18]. Analogous to the temperature induced electronic transition, the adjustability in the hydrogen triggered electronic transition properties of ReNiO$_3$ is expected to be largely expanded via further extending the investigations towards heavier rare-earth composition beyond Eu.

In this work, we demonstrate the fundamental relationship between the rare-earth composition and the hydrogen induced electronic transition properties for quasi-single crystalline ReNiO$_3$/LaAlO$_3$ (001) heterostructures, covering a large variety of the rare-earth composition from Nd to Er. Assisted by the near edge X-ray absorption fine structure (NEXAFS) analysis, the transition in electronic structures and orbital configurations of ReNiO$_3$ upon hydrogenation are demonstrated. A non-monotonic tendency in the hydrogenation elevated material resistivity versus the $r_{Re}$ or the atomic number of the rare-earth composition is discovered, owing to two different competing mechanisms. Balancing the hydrogen induced transition reversibility with the abruption in the variations of material resistivity, we emphasize that the ReNiO$_3$ with middle rare-earth compositions (e.g. Sm) to be most suitable that caters for the potential applications in correlated electronic devices.

The spontaneous electronic transition within ReNiO$_3$ induced by hydrogen was demonstrated to be thermodynamically driven by the material metastability at the Ni$^{3+}$ based $t_{2g}^6 e_g^{1\pm\Delta}$ (or $t_{2g}^6 e_g^1$) orbital configuration[11,19,20]. As illustrated in Figure 1a, the hydrogenation reconfigures the electronic structure of ReNiO$_3$ towards electron localized state of Ni$^{2+}$ based $t_{2g}^6 e_g^2$, the process of which meanwhile reduces the positive magnitude in $\Delta G$. It is worth noticing that reducing $r_{Re}$ elevates the $\Delta G$ of ReNiO$_3$, the hydrogen induced electronic transitions is expected to be intrinsically dominated by the rare-earth composition.

In Figure 1b and 1c, we show the thermodynamic phase stability of ReNiO$_3$ (e.g. Re=Nd, Sm, Gd and Dy) in the pressure to temperature plane (P–T) as predicted according to ref[21]. The dash line represents the merit for $\Delta G$=0 for the heterogeneous growth of ReNiO$_3$, as calculated by the following equation:

$$\Delta G = \Delta H_{LNO,1000K} - T\Delta S_{LNO,1000K} + h(r(Re^{3+}) - r(La^{3+})) - (1/4)RT \ln(P) \quad ,\quad \text{where}$$

$\Delta H_{LNO,1000K}$ and $\Delta S_{LNO,1000K}$ are the enthalpy and entropy changes of LaNiO$_3$ at 1000 K, $h$ and $s$ are constants taken from ref[21], $r(Re^{3+})$ and $r(La^{3+})$ are the ion radius of $Re^{3+}$ and $La^{3+}$, and $R$ is the ideal gas constant. Nevertheless, the coherent relationship between ReNiO$_3$ and the LaAlO$_3$ (001) substrate is expected to stabilize the metastable phase via interfacial chemical bonds (e.g. -0.23 eV/Å$^2$)[22], and the respective solid line downwards demonstrate the merit for $\Delta G$=0 by considering such heterogeneous interfacial effect. It can be seen that ReNiO$_3$ with heavier rare-earth composition requires higher magnitude of oxygen partial pressure to stabilize the distorted perovskite structure. As more clearly demonstrated in Figure 1d, at a representative hydrogenation temperature as previously reported (e.g. 100 °C), the $\Delta G$=0 is more positive for DyNiO$_3$ and GdNiO$_3$, compared to SmNiO$_3$ and NdNiO$_3$. Therefore, from a thermodynamic perspective, this shed a light on further improving the hydrogen induced electronic transition properties for using ReNiO$_3$ with heavier rare-earth composition.

To explore how the rare-earth composition impacts the hydrogen induced electronic transition properties of ReNiO$_3$, we grew a serial of ReNiO$_3$/LaAlO$_3$ samples with various Re compositions, including Nd, Sm, Gd, Dy, Y and Er according to our previous reports[23]. The resistances (R) of as-grown films were measured as a function of temperature (T), and their respective ln$R$-1000/$T$ tendencies are compared in Figure 2a. The temperature dependence in electronic transportations varies abruptly for NdNiO$_3$, SmNiO$_3$ and GdNiO$_3$ thin film samples, when elevating (or descending) temperature across the $T_{MIT}$. In contrast, the $T_{MIT}$ for DyNiO$_3$, YNiO$_3$ and ErNiO$_3$ were not measurable, since the metastable thin film samples decomposed when elevating the temperature of characterizations above, e.g. 260 °C, in the Ar atmosphere before reaching their transition temperatures. Nevertheless, the formations of their metastable perovskite phase are demonstrated by the thermistor transportation behavior with large magnitude of negative temperature coefficient resistance (NTCR) in their insulating phase.

According to the previous reports, the insulating phase of $Re$NiO$_3$ follows a hopping principle in their electrical transportation and the respective thermal activation energy ($E_a$) can be estimated from their respective ln$R$-1000/$T$ tendencies[24]. In Figure 2b, the magnitude of $E_a$ is further compared for $Re$NiO$_3$/LaAlO$_3$, while more details of the fitting are shown in Figure S8 and S9. It can be seen that reducing the $r_{Re}$ enlarges the magnitude of $E_a$, in particular, near room temperature. This result is in agreement to the increasing tendency in the band gap of $Re$NiO$_3$ with heavier rare-earth composition[12,25], and thereby the carrier (electron) hopping requires to overcome larger energy barrier.

To trigger the electronic transition via hydrogen, dot-shaped platinum pattern was grown on the surface of the $Re$NiO$_3$/LaAlO$_3$ (see structure illustration in Figure S10), and afterwards annealed in 20% H$_2$/Ar mixing gas at 100 °C for 30 minutes. In Figure 3a, the resistances of Pt/$Re$NiO$_3$/LaAlO$_3$ as measured between the two adjacent platinum dots are compared before and after the hydrogenation process. Prior to the hydrogenation process, the initial resistance ($R_0$) of $Re$NiO$_3$ shows an increasing tendency with a reducing $r_{Re}$, and this is in agreement to the elevation in the material resistivity for heavier rare-earth composition. The following hydrogenation process significantly enlarges the resistance ($R_H$), in particular, for $Re$NiO$_3$ with rare-earth composition lighter than Dy. In Figure 3b, the relative elevation in resistance upon hydrogenation ($R_H$/$R_0$) is further compared. It is worth noticing that the $R_H$/$R_0$ firstly shows an increasing tendency with a reducing $r_{Re}$ until Dy. Nevertheless, by further reducing $r_{Re}$ towards the heavier rare-earth composition (e.g. Y, Ho and Er), the $R_H$/$R_0$ shows a reducing tendency.

The above non-monotonic $R_H$/$R_0$-$r_{Re}$ tendency unveils two competing mechanisms in the regulations of the hydrogen induced electronic transition properties of $Re$NiO$_3$ via the rare-earth compositions, as illustrated in Figure 3c. As already pointed out previously, reducing the $r_{Re}$ is expected to enlarge the positive magnitude in $\Delta G$ of $Re$NiO$_3$, and this provides larger driving force of the hydrogenation that promotes the resultant electronic transition. As a result, the magnitude in $R_H$/$R_0$ is elevated by reducing $r_{Re}$ from Nd towards Dy. Nevertheless, the initial material resistivity of $Re$NiO$_3$ will be increased by reducing $r_{Re}$, while the metastable phase of $Re$NiO$_3$ with heavy rare-earth composition (e.g. Dy, Y or Er) is more difficult to be stabilized. It results in the decomposition of these $Re$NiO$_3$ with heavy rare-earth composition during the hydrogenation process (e.g. into NiO$_x$ and $Re$O$_x$), instead of triggering the formation of the expected electron localized electronic phase related to H-$Re$NiO$_3$.

This explains the afterwards reducing tendency in $R_H/R_0$ by further reducing $r_{Re}$ from Dy towards Er.

The above understanding is further confirmed by the reversibility in the hydrogenation induced electronic transition of $Re$NiO$_3$, e.g., by further performing dehydrogenation process to anneal the hydrogenated samples in air at 300 °C for 60 minutes. As shown in Figure 3d, the resistance for $Re$NiO$_3$ with lighter rare-earth composition (e.g. Sm) recovers to the pristine magnitude upon dehydrogenation. This is in contrast to $Re$NiO$_3$ with heavy rare-earth composition (e.g. Gd or Dy), in which situation the resistance partially recovered by performing the same hydrogenation and dehydrogenation process.

To further demonstrate the electronic structures of $Re$NiO$_3$ with various rare-earth compositions during the hydrogen induced electronic transition, the near edge X-ray absorption fine structure (NEXAFS) analysis was performed. Figure 4 shows the NEXAFS spectra of the Ni-$L$ edges and O-$K$ edges for NdNiO$_3$ (Figure 4a and 4b) and SmNiO$_3$ (Figure 4c and 4d) before and after the hydrogenation process. The NEXAFS spectrum of the Ni-$L_3$ edge reflects the Ni 2p→Ni 3d transition and splits into peak A and B, while the Ni-$L_2$ edge is associated to Ni-O hybridization strength[26]. The O-$K$ edge is split into pre-peak A and peaks B and C, where the pre-peak A is attributed to the Ni$^{3+}$ ($d^8L$) configuration, while peaks B and C are associated to the Ni$^{2+}$ ($d^9L$) configuration.

A reduction in the relative intensity of the peak B split from peak A within the Ni-$L_3$ edges is observed for both SmNiO$_3$ and NdNiO$_3$ upon hydrogenation. This demonstrates the decrease in the proportion of $t^6_{2g}e^1_g$ (Ni$^{3+}$) orbital configuration compared to the $t^6_{2g}e^2_g$ (Ni$^{2+}$)[27], indicating a strengthened electron localized electronic phase compared to the electron itinerant one. Further accordant tendency was observed in their NEXAFS spectra of O-$K$ edges, in which case a reduction in the peak A of the O-$K$ edge is observed upon hydrogenation. This indicates a reduced oxygen-projected density of unoccupied electronic states owing to hydrogen induced electron filling, and is in agreement to the increasing proportion in the Ni$^{2+}$ orbital configuration compared to Ni$^{3+}$.

In summary, we demonstrate the relationship between the rare-earth composition and the hydrogen induced electronic phase transition properties for quasi-single crystalline $Re$NiO$_3$/LaAlO$_3$ (001) heterostructures, covering a large variety of the rare-earth composition from Nd to Er. Upon the same hydrogenation process, the magnitude in $R_H/R_0$ firstly shows an increasing tendency from $10^3$ to $10^7$ by reducing the $r_{Re}$ from Nd to Dy, and afterwards reduces to $10^2$ for further reducing $r_{Re}$ towards Er. The above unexpected non-monotonic $R_H/R_0$-$r_{Re}$ tendency reveals the following two competing mechanisms that dominate the hydrogen induced electronic phase properties of $Re$NiO$_3$ with heavy rare-earth compositions. On the one hand, the larger positive $\Delta G$ when reducing $r_{Re}$ promotes the driving force of the hydrogenation process from the thermodynamic perspective, and this is expected to enlarge $R_H/R_0$. On the other hand, the $Re$NiO$_3$ with smaller $r_{Re}$ exhibits higher initial resistivity and is easier to be decomposed during the hydrogenation process, both of which reduces the $R_H/R_0$. Further probing the electronic structures via near edge X-ray absorption fine structure analysis clearly demonstrates the respective transition in electronic structures of $Re$NiO$_3$ from Ni$^{3+}$ based electron itinerant orbital configurations towards the Ni$^{2+}$ based electron localized state. Balancing the hydrogen induced transition reversibility with the abruption in the variations of material resistivity, we emphasize that the $Re$NiO$_3$ with middle rare-earth compositions (e.g. Sm) to be most suitable that caters for the potential applications in correlated electronic devices.

**Supplementary material**

Supplementary material is available online. Figure S1-S7: $U$-$I$ of $Re$NiO$_3$ and H-$Re$NiO$_3$; Figure S8 and S9: ln$\rho$-1000/$T$; Figure S10: Pt/$Re$NiO$_3$/LaAlO$_3$.

**Acknowledgments**

This work was supported by the National Key Research and Development Program of China (No. 2021YFA0718900), National Natural Science Foundation of China (Grant No. 62074014 and 52073090). In addition, JC also acknowledge the support by Beijing New-star Plan of Science and Technology (No. Z191100001119071).

**Competing interests**

We declare no competing financial interest.

**Additional information:** Supplementary Information is available for this manuscript.

**Correspondence:** Prof. Jikun Chen (jikunchen@ustb.edu.cn).

**Figures and captions**

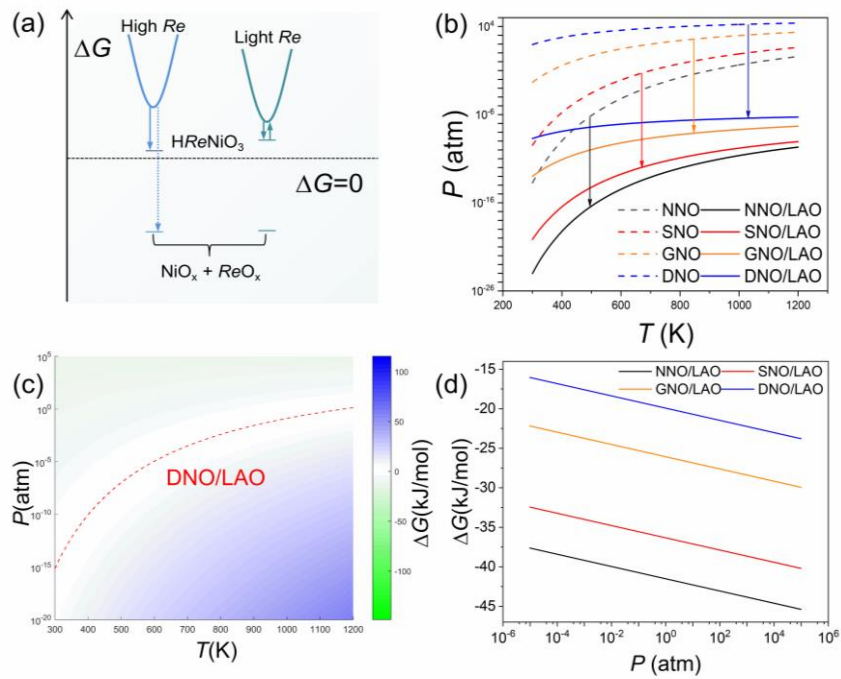

**FIG. 1.** (a) Illustration of the Gibbs free energy change (ΔG) of *Re*NiO$_3$ before and after the hydrogenation process. (b) Thermodynamic phase stability diagram for *Re*NiO$_3$ (*Re*=Nd,Sm,Gd and Dy) with interfacial reaction. (c) Temperature dependence of the pressure (left: P-T) and the Gibbs free energy change (right: ΔG-T) for DyNiO$_3$ (DNO) on LaAlO$_3$ (LAO). (d) Plot of ΔG vs P for *Re*NiO$_3$/LAO.

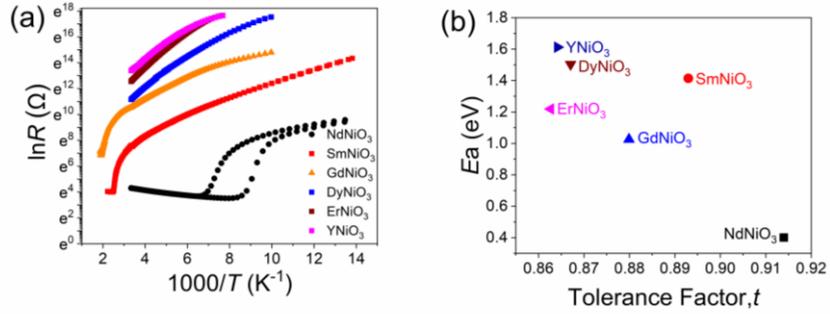

**FIG. 2.** (a) Temperature dependence of resistivity (lnR) for $Re$NiO$_3$ ($Re$=Nd,Sm,Gd,Dy,Er and Y). (b) The thermal activation energy (E$_a$) of $Re$NiO$_3$.

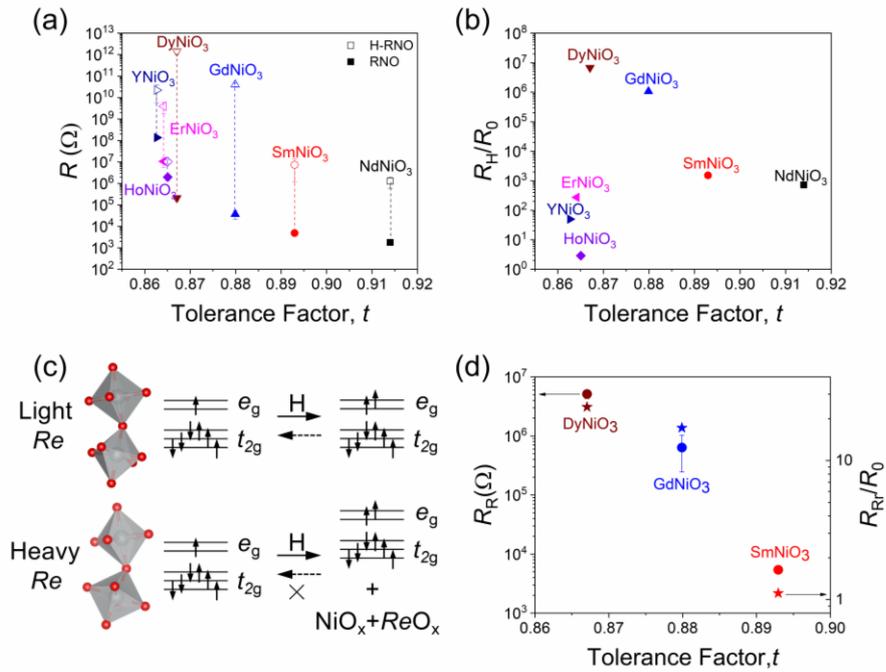

**FIG. 3.** (a) The resistances of $Re$NiO$_3$ before and after hydrogenation. (b) The ratio of the hydrogenation resistance and initial resistance (R$_H$/R$_0$) of $Re$NiO$_3$. (c) Illustration of two competing mechanisms in the regulations of the hydrogen induced electronic transition properties of $Re$NiO$_3$ via the rare-earth compositions. (d) The recovered resistance(R$_R$) and the ratio of the recovered resistance and initial resistance (R$_R$/R$_0$) of SmNiO$_3$, GdNiO$_3$ and DyNiO$_3$.

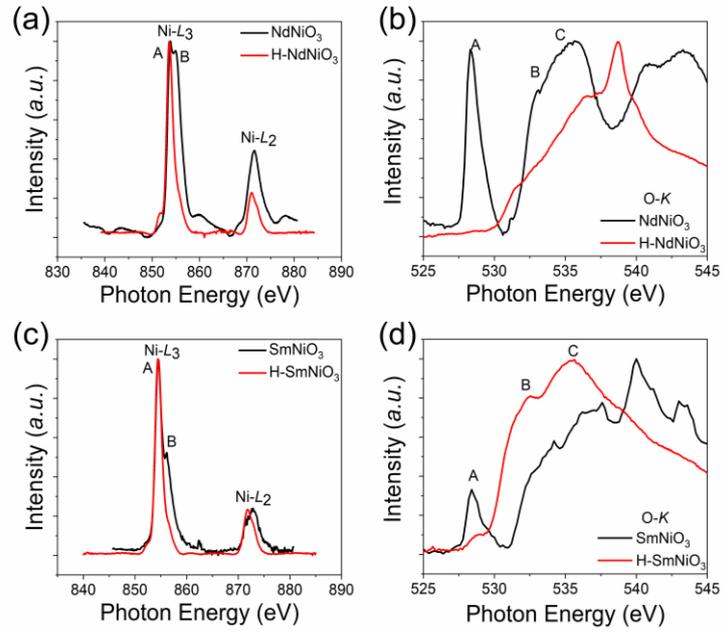

**FIG. 4.** The NEXAFS spectra of the Ni-L edges and O-K edges for (a) and (b) NdNiO$_3$, (c) and (d) SmNiO$_3$.